\documentclass[3p,times,procedia]{elsarticle}
\usepackage{nupha_ecrc}
\biboptions{sort&compress}
\usepackage{wrapfig}
\usepackage{amsmath}

\volume{00}

\firstpage{1}

\journalname{Nuclear Physics A}

\runauth{}

\jid{nupha}

\jnltitlelogo{Nuclear Physics A}

\usepackage{amssymb}

\usepackage[figuresright]{rotating}

\begin{document}

\begin{frontmatter}



\dochead{XXVIIIth International Conference on Ultrarelativistic Nucleus-Nucleus Collisions\\ (Quark Matter 2019)}

\title{Quarkonium production and suppression: Theory}


\author{Alexander Rothkopf}

\address{Faculty of Science and Technology, University of Stavanger, 4021 Stavanger, Norway}

\begin{abstract}
Heavy quarkonium theory has seen significant progress over the past years. Not only has the community steadily improved the understanding of fully equilibrated quarkonium using first principles lattice QCD studies, but it has taken major steps towards developing a genuine dynamical understanding of quark-antiquark pairs immersed in a hot medium by embracing the paradigm of open quantum systems. We review recent developments, placing an emphasis on quarkonium real-time dynamics.
\end{abstract}


\end{frontmatter}



\vspace{0.75cm}

Heavy quarkonium is an exceptionally versatile laboratory to study the physics of the strong interactions in extreme conditions (for a recent review see \cite{Rothkopf:2019ipj}) and possesses many favorable properties from the point of view of both the experimental and theory community. One advantage is that heavy quark-antiquark pairs decay predominantly into dilepton pairs, producing clean experimental signals in relativistic heavy-ion collision. This has allowed the precision determination of their production abundances (see e.g. the CMS bottomonium di-muon spectra \cite{Sirunyan:2017lzi}), as well as their elliptic flow (see e.g. the ALICE charmonium $v_2$ and $v_3$ \cite{Acharya:2018pjd}). In the eyes of theory, quarkonium constitutes one of the simplest strongly interacting systems. The inherent separation of scales between the heavy quark rest mass $m_Q$ and other relevant scales in the system, such as the intrinsic scale of quantum fluctuations in Quantum Chromo Dynamics (QCD) $\Lambda_{\rm QCD}$ or the environment energy density $\varepsilon_{\rm env}$ enables the use of powerful so-called effective field theory (EFT) methods. This in turn allows theory to deploy an intuitive non-relativistic language to describe quarkonium in a hot environment in terms of e.g. an in-medium potential (see e.g. \cite{Laine:2006ns,Rothkopf:2011db,Burnier:2014ssa,Petreczky:2018xuh}) or transport coefficients, such as the heavy-quark diffusion constant (see e.g. \cite{Kaczmarek:2020umv, Brambilla:2019tpt}). Combining the insight on quarkonium from experiment and theory, our goal is two-fold: we wish to gain a quantitative understanding of QCD bound states in a hot environment and exploit this knowledge to shed light on the properties of nuclear matter created in heavy-ion collisions.

To fully understand heavy quarkonium in a heavy-ion collision, one must recognize that its life cycle encompasses all stages of such a collision. The quark-antiquark pair is produced in the very early stages and needs to shed its color through gluon emission before it can form a quarkonium particle, if at all. Theory exploits the fact that the production of the heavy quark pair takes place at a high enough energy scale so that perturbation theory is applicable (nb.\ non-perturbative nuclear parton distribution functions however play an important role here too). In addition $Q\bar{Q}$ production factorizes from the quarkonium formation process, which itself is genuinely non-perturbative. Theory is making progress in quantitatively elucidating quarkonium production in p+p and p+A collisions by improving on phenomenological models, such as the color evaporation model \cite{Cheung:2018upe} and by recently marrying the effective field theory NRQCD and the color glass condensate framework \cite{Ma:2015sia}. A similarly robust understanding of early production and formation in A+A has however not yet been established.

If a quarkonium state forms early on, it finds itself soon immersed in the quark-gluon-plasma formed by the light remnants of the collision. The incessant kicks of this hot environment will interfere with the binding of the quark-antiquark pair, which may either survive or dissociate along the way. Once the bulk matter reaches the crossover transition to the hadronic phase, colored partons have to combine into color neutral hadrons and we can entertain two possible scenarios: either a surviving in-medium state transitions into a vacuum state or (as we now know is relevant for charmonium at LHC) if a large enough number of quark-antiquark pairs was produced early on, there exists a finite probability for them to recombine into quarkonium even if they became fully dissociated in the QGP. Theory is actively exploring the consequences of a scenario in which all quarkonium is produced at the phase transition based on the statistical model of hadronization \cite{Andronic:2019wva} and at this conferences new ideas on the coalescence of single heavy quarks into open heavy flavor during hadronization have been presented \cite{He:2020umv,Zhao:2020wpi}.
\begin{figure}
\centering
\includegraphics[scale=0.4]{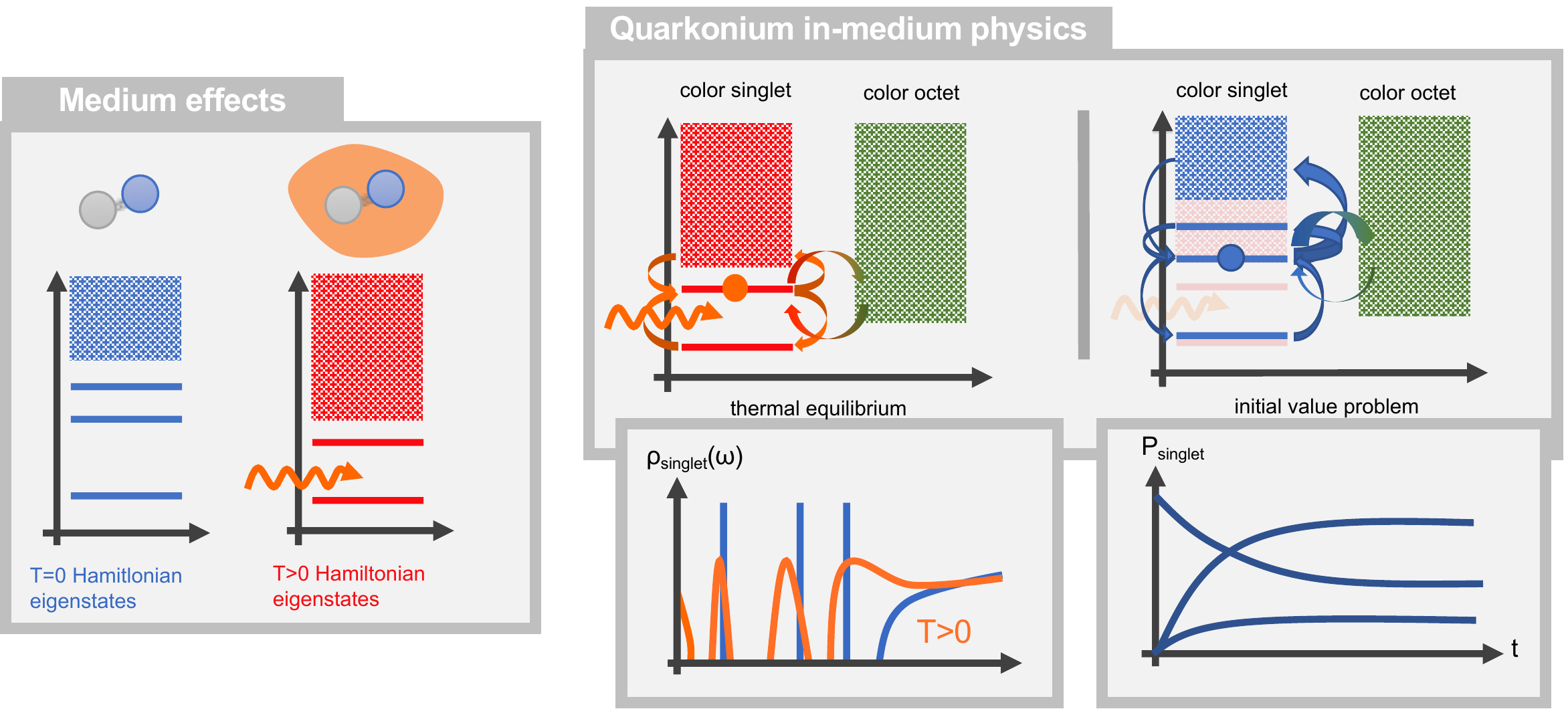}
\caption{Overview of the different physics mechanism at play for in-medium heavy quarkonium}\label{Fig:overview}
\vspace{-0.5cm}
\end{figure}

In the following let me concentrate on the recent progress made in understanding heavy quarkonium in the intermediate stages of the collision, where one considers a fully formed state ineracting strongly with its hot environment. To get an intuition for the physics involved, I invite the reader to think of heavy-quarkonium in terms of an energy level diagram, often used in atomic physics and as shown on the left of Fig.\ref{Fig:overview}.\ Already Matsui and Satz \cite{Matsui:1986dk} knew that in the presence of a medium the Hamiltonian of the quarkonium system changes, i.e. some former bound stationary states become unbound and the energy of the remaining ones may change (blue to red levels). What has been understood however over the past 12 years is that the medium also acts as a radiation field (orange arrow) which may introduce absorption and emission processes. Imagine we start with an color-singlet eigenstate of the in-medium Hamiltonian (orange dot, center box). Instead of remaining set, the medium will induce transitions to higher or lower lying states, as well as between the singlet and octet sector. In thermal equilibrium the transitions among states are balanced out, such that the probability to find the particle at a certain energy remains time-independent. Theory has non-perturbative access to this physics via the study of in-medium equilibrium spectral functions $\rho(\omega)$ from e.g. lattice QCD. At $T=0$ they contain delta-like peaks where bound states exist (plus a continuum) while at $T>0$ the peaks broaden and shift. The latter indicates the change in the Hamiltonian, the former encodes the probability to transition away from the state. In a heavy-ion collision the situation is different, as one starts out with an unequilibrated state (blue dot, top left figure). Now both the Hamiltonian induces mixing between the states, as well as the radiation field and these transitions are not balanced. Only at late times will the system equilibrate to time independent occupations. Here we are interested in the real-time evolution of the occupation numbers, or what is conventionally termed survival probabilities of different states. A lot of progress has been made to capture this physics, but a fully non-perturbative approach is still outstanding.

Extracting equilibrium spectral functions from lattice QCD is challenging, as it amounts to a highly ill-posed inverse problem. A lattice QCD simulation acts similarly to an imperfect detector producing a spectrum that is folded with a resolution function. Actually lattice QCD encodes $\rho(\omega)$ in so-called Euclidean-time correlation functions $D(\tau)$, which are essentially their Laplace transform. Two strategies exist on the market to extract the spectral function: on the one hand we may use Bayesian inference, a data analysis technique which exploits prior QCD domain knowledge (e.g. positivity of  $\rho(\omega)$) to regularize the inverse Laplace transform. The well known Maximum Entropy Method (MEM) \cite{Asakawa:2000tr} or the Bayesian Reconstruction (BR) method \cite{Burnier:2013nla} are two different incarnations of this strategy. While they do not presuppose a particular form of the spectral function they require very high precision input data to succeed. On the other hand one may model the spectral function based on prior information obtained from EFT's and perturbation theory and then fit the model parameters to the lattice data. This strategy gives access to e.g. excited state and transport physics but introduces model dependencies.

The Bayesian inference community has made lot's of progress understanding the systematic effects of the reconstruction process over the past years and is converging on consistent results among different groups e.g. for the stability estimates of ground state quarkonium \cite{Aarts:2014cda,Kim:2018yhk}. While recent studies based on non-relativistic lattice EFT's have provided consistent results indicating that the mass shifts of quarkonium are negative, it remains an active research question. Recent results on $T>0$ quarkonium dispersion relations were presented in \cite{Ikeda:2016czj}. All Bayesian studies so far agree that in-medium modification occurs in a sequential fashion, related to the vacuum binding energy of the quarkonium state. It needs to be kept in mind that this of course does not directly translate into sequential suppression in a heavy-ion collision. Limited by simulation data precision, so far no results on excited states and only limited information of spectral widths has been obtained.

Two quarkonium spectral modelling studies in lattice QCD have been presented at this conference. In \cite{Larsen:2020umv} a non-relativistic formulation of heavy quarks is deployed on realistic simulations ($m_\pi \approx 160$MeV) of the thermal QCD medium background. Adopting techniques from $T=0$ simulations, i.e. the use of extended sources for a better signal to noise ratio, to $T>0$ and utilizing a simple parametrization of the spectral function, the authors fit the lattice data to obtain first results on the in-medium modification of bottomonium excited states and the width of the ground state. Again hierarchical ordering with vacuum binding energy is observed. The second study \cite{Kaczmarek:2020umv} considers fully continuum extrapolated simulation data based on a relativistic formulation of heavy quarks, which however are immersed in a purely gluonic medium only. The spectral functions here are modelled using combined prior information from perturbation theory and the EFT pNRQCD. Fits of the lattice data reveal a negative mass shift and allow to give a robust lower limit on the transport coefficient related to heavy quark diffusion $D>2/2\pi T$. 

Besides obtaining spectral information from lattice QCD, the definition and extraction of an intuitive in-medium heavy quark potential has been one focus within heavy quarkonium theory. We now understand how to define it systematically from QCD by approximating the physics of heavy quarks via a set of EFT's called NRQCD and pNRQCD (see e.g \cite{Brambilla:2004jw}). In the latter, quarkonium is represented as color singlet and octet wavefunctions and the Lagrangian of these degrees of freedom directly reflects the physical processes discussed in the center panel of Fig.\ref{Fig:overview}. The in-medium energy levels are given by a Schr\"odinger-like part in the Lagrangian, where the singlet and octet evolve according to their respective in-medium potentials. In addition radiation field effects of the gluons are included by dipole-like terms, which may induce transitions between the two sectors. 

The parameters of this EFT, the interactions potentials and the strength of the dipole terms are set via a matching procedure, where we compute a correlation function in the EFT and a correlation function with the same physics content in QCD and set them equal at the energy scale of interest. In the limit of static quarks the so-called real-time Wilson loop is the QCD counterpart to the pNRQCD correlation function of two color-singlet wavefunctions (see e.g. \cite{Laine:2006ns,Beraudo:2007ky,Brambilla:2008cx}). Thus their real-time evolution can be identified and used to unambiguously define what we mean by the in-medium heavy quark potential in the language of EFTs. Note that this quantity in general can be complex. The real-time definition cannot directly be evaluated in lattice QCD simulations, which are carried out in an artificial Euclidean time but strategies have been developed \cite{Rothkopf:2011db,Burnier:2012az,Burnier:2014ssa}, based on spectral function reconstruction, to extract its values non-perturbatively. The most recent results on the real- and imaginary part of the potential, based on Bayesian spectral reconstructions and a Pade approximation technique were given in \cite{Petreczky:2018xuh}. This real-time potential has also been computed using methods of holography, presented as poster at this conference (see e.g. \cite{Barnard:2019mnk}).

Other definitions of potentials exist in the literature, which are not based on the above mentioned EFT's. E.g. the T-matrix approach \cite{Liu:2017qah} sets out to describe the physics of both the light and heavy quark degrees of freedom with a model Hamiltonian, which includes a kinetic term, as well as a four-fermi interaction, whose strength is given by a real-valued in-medium potential. The functional form of this potential is fitted, such that the model reproduces thermodynamic properties of QCD computed on the lattice, e.g. the QCD free energies. Remarkable agreement with the non-perturbative data is found. Since however the T-matrix potential is not based on a separation of scales and attempts to summarize the the physics of light and heavy quarks it is conceptually different from the EFT potential and thus it is neither surprising nor of concern that the best fit values for the T-matrix potential differ from the lattice EFT potential. 

The study of the heavy quark potential in lattice QCD is one way how to build a first-principles bridge from QCD at finite temperature to the real-time dynamics of heavy quarkonium. First-principles insight is indeed needed, as the currently available experimental data (mostly $R_{AA}$ of quarkonium ground state particles) can be reproduced by a variety of phenomenological models, which incorporate widely different physics mechanisms. We can distinguish two large classes of studies: one uses a rate equation where a temperature dependent decay rate drives the system towards an equilibrium distribution (see e.g. \cite{Ferreiro:2012rq,Zhao:2011cv,Liu:2009nb}). The other, deployed mainly for bottomonium, solves a deterministic Schr\"odinger equation in the presence of a complex in-medium potential (see e.g. \cite{Krouppa:2016jcl,Krouppa:2017jlg}). If we wish to quantitatively infer the physics underlying the experimental results and distinguish between different models, we need to establish their actual range of validity and interrogate QCD about appropriate parameters to use in these models. In addition, theory has the goal to derive more and more accurate real-time descriptions of quarkonium directly from QCD.

An important step towards establishing a general real-time approach to heavy quarkonium has been achieved in the community by embracing the open quantum systems framework. Originally developed in the context of condensed matter theory, it focusses on scenarios in which a small subsystem is immersed in a large (thermal) environment. The overall system, consisting of environment and subsystem, is closed. It possesses a hermitean Hamiltonian, which itself is composed of a part describing the interactions among the quark-antiquark pair $H_{Q\bar{Q}}$, another part for the medium degrees of freedom $H_{\rm med}$ and the interaction part between the two (the latter is often written as $H_{\rm int}=\sum_m \Sigma_m\otimes \Xi_m$, the former acting in the quarkonium subspace, the latter in the medium subspace). The overall density matrix $\rho(t)$ evolves according to the von-Neumann equation\vspace{-0.9cm}

\begin{align}
H=H_{Q\bar{Q}}\otimes I_{\rm med} + I_{Q\bar{Q}} \otimes H_{\rm med} + H_{\rm int}, \quad \frac{d\rho}{dt}=-i[H,\rho].
\end{align}\vspace{-0.5cm}

\noindent What we are interested in however is only the dynamics of the small $Q\bar{Q}$ subsystem. Formally tracing out the medium degrees of freedom leads to the so-called reduced density matrix $\rho_{Q\bar{Q}}={\rm Tr}_{\rm med}[\rho]$. The goal of theory is to derive from QCD the equation of motion for $\rho_{Q\bar{Q}}$, called the master equation. A generic feature is that tracing out the medium leads to dissipative dynamics of the subsystem, even if the evolution of the overall system is fully unitary. The form of these master equations depends on the particular separation between three characteristic time-scales: The {\it environment relaxation scale} $\tau_E$ from $\langle \Sigma_m(t)\Sigma_m(0)\rangle\sim e^{-t/\tau_E}$, the {\it intrinsic $Q\bar{Q}$ scale} $\tau_S$ from $\tau_S\sim1/|\omega-\omega^\prime |$ and the {\it $Q\bar{Q}$ relaxation scale} $\tau_{\rm rel}$, which for a heavy particle can be defined from the isotropization timescale of its momentum $\langle p(t)\rangle \sim e^{-t/\tau_{\rm rel}}$. In case that the environment relaxes much faster than the quarkonium system, the time evolution can be approximated as Markovian. This in turn allows it to be cast in the general form of a {\it Lindblad equation}\vspace{-0.5cm}

\begin{align}
\frac{d}{dt}\rho_{Q\bar{Q}}=-i\big[\tilde H_{Q\bar{Q}},\rho_{Q\bar{Q}}\big]+\sum_k\gamma_k\Big(L_k\rho_{Q\bar{Q}}L^\dagger_k -\frac{1}{2}L^\dagger_kL_k \rho_{Q\bar{Q}}+\frac{1}{2}\rho_{Q\bar{Q}}L_k^\dagger L_k\Big)\label{Eq:Lindblad},
\end{align}\vspace{-0.5cm}

\noindent which conserves hermiticity $\rho_{Q\bar{Q}}^\dagger=\rho_{Q\bar{Q}}$, the probability interpretation $Tr[\rho_{Q\bar{Q}}]=1$ and positivity $\langle n|\rho{Q\bar{Q}}|n\rangle >0, \forall n$ during the time evolution. The modified Hamiltonian $\tilde H_{Q\bar{Q}}$ again encodes the energy level changes of Fig.\ref{Fig:overview}, while the Lindblad operators $L_k$ are intimately related to the radiation field effects of the medium.

\begin{figure}
\centering
\includegraphics[scale=0.5]{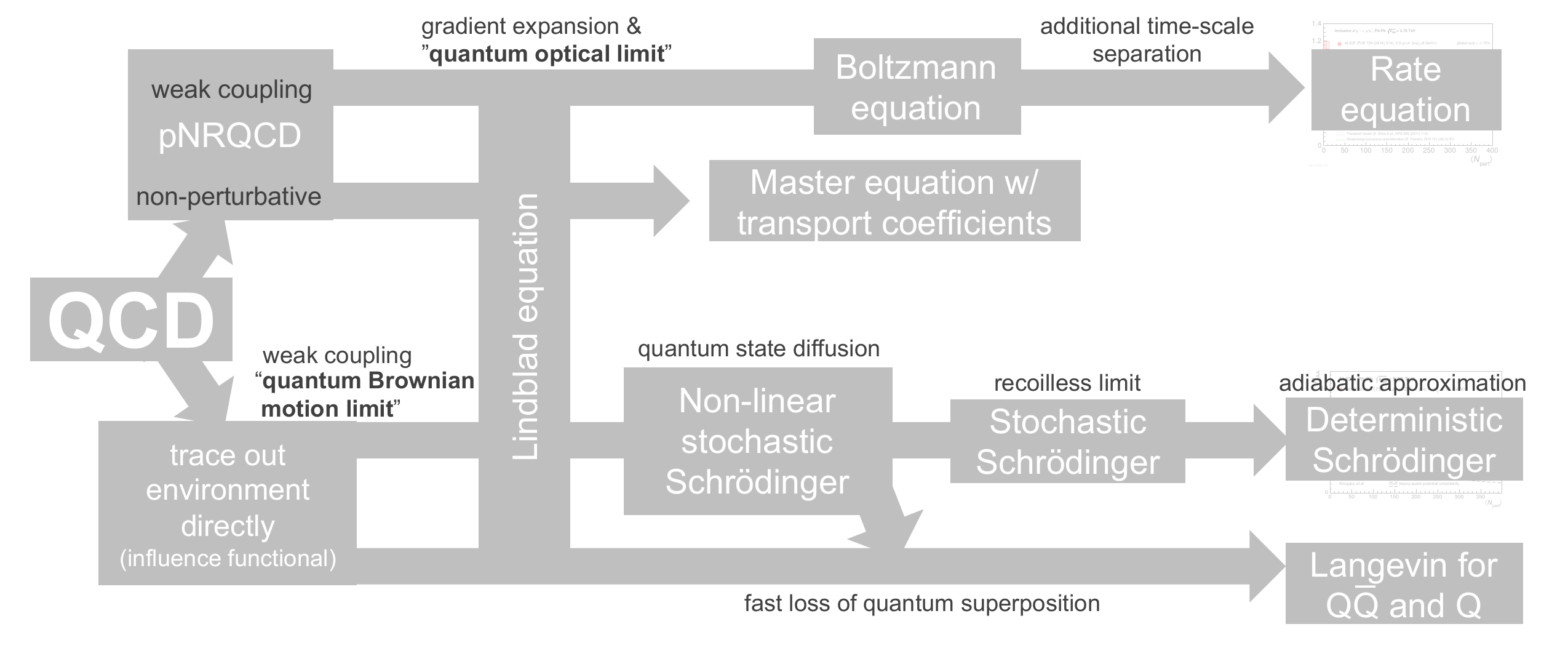}\vspace{-0.3cm}
\caption{Overview the current open-quantum-system based approaches to quarkonium dynamics. Note that they provide a systematic path from QCD to the most popular phenomenological models on the market through a series of well specified approximations, establishing robustly a range of validity.}\label{Fig:OQSapproaches}
\vspace{-0.5cm}
\end{figure}

As shown in Fig.\ref{Fig:OQSapproaches} two major branches are being explored in this line of study at the moment. On the one hand theorists use their experience with the EFT pNRQCD at high temperature to derive a Lindblad equation applicable for a weakly coupled medium. Assuming the time scale separation of the quantum optical limit $\tau_S\ll\tau_{\rm rel}$ and performing a gradient expansion it has been shown how to reduce that Lindblad equation to the Boltzmann equation and in case of further time-scale separations even further to a rate equation. As presented at this conference \cite{Yao:2020kqy} this result was achieved by expressing the operators $\Sigma$ and $\Xi$ of the interaction Hamiltonian by the quantities present in the weak-coupling pNRQCD Lagrangian and relating the distribution function of the Boltzmann equation to the color singlet matrix elements of the reduced density matrix \cite{Yao:2018nmy}. The important point here is that this derivation provides consistent pNRQCD based expressions for the recombination and dissociation scattering kernels in the Boltzmann equation. Simplifying the description of the color octet distribution functions by treating dissociated quarks and antiquarks via a Langevin diffusion equation, this Boltzmann equation has been deployed to estimate quarkonium yields in \cite{Yao:2018sgn}.

In order to proceed towards a non-perturbative treatment, another group has focused on the subset of deep Coulombically bound quarkonium states immersed in a strongly coupled medium \cite{Brambilla:2019tpt}. It can be shown that in this specific case the dynamics are governed by two transport coefficients, the heavy quark diffusion constant, as well as a potential correction term. As they can be expressed in terms of correlation functions of color electric fields, these transport coefficients are amenable to an evaluation in lattice QCD, a task whose completion is ongoing. As is the case with all pNRQCD based computations, we must keep in mind that it is formulated for a specific hierarchy of scales, which needs to be ascertained when deploying these methods in a phenomenological simulation concurrently evolving many different quarkonium states. 

Along the second, lower branch in Fig.\ref{Fig:OQSapproaches} the community works on explicitly tracing out the medium degrees of freedom in the path integral language. This leads to the so-called Feynman-Vernon influence functional, which encodes all information about $Q\bar{Q}$ medium interactions. In general it represents an extremely complicated expression and only through further approximations can it be tamed, giving way to a Markovian Lindblad equation. Interestingly it is possible to show that such master equations are equivalent to a non-linear stochastic Schr\"odinger equation. This has allowed for the first time to give a QCD derivation to such non-linear approaches (see e.g. \cite{Katz:2015qja}) previously introduced in the phenomenology community. Going to the recoilless limit (neglecting dissipation) the dynamics can be reduced to a linear Schr\"odinger equation, which recovers another phenomenological model, the stochastic potential of Ref.\cite{Akamatsu:2011se}. Finally going to the adiabatic limit one recovers the deterministic Schr\"odinger equation currently deployed in the study of Bottomonium states (see e.g. \cite{Krouppa:2016jcl}) . Again we are able to give a {\it systematic chain of approximations} and thus clear ranges of validity for phenomenological models thanks to the open quantum systems approach.

Let us focus on one strategy to derive the master equation, which focuses on high temperatures. There, the medium is weakly coupled and thus a perturbative expansion may be deployed. Assuming the time-scale separation of the quantum-Brownian motion limit $\tau_E\ll \tau_{S}$, which is often realized among quarkonium states, the following momentum and color dependent Lindblad operator has been derived in Ref. \cite{Akamatsu:2012vt,Akamatsu:2014qsa}\vspace{-0.3cm}

\begin{align}
\nonumber L_{{\bf k},a}=\sqrt{\frac{D({\bf k})}{2}} \Big[1-\frac{{\bf k}}{4m_QT}\Big(\frac{1}{2}{\bf P}_{\rm CM}+{\bf p}\Big)\Big]e^{i{\bf k}{\bf r}/2}(T^a\otimes 1)
-\sqrt{\frac{D({\bf k})}{2}} \Big[1-\frac{{\bf k}}{4m_QT}\Big(\frac{1}{2}{\bf P}_{\rm CM}+{\bf p}\Big)\Big]e^{-i{\bf k}{\bf r}/2}(1\otimes T^a).\vspace{-0.3cm}
\end{align}
I emphasize this equation, since it embodies an interesting connection between quarkonium real-time dynamics and the EFT based heavy quark potential discussed before. The two terms correspond to the physics of the quark and antiquark respectively (overall color neutrality of quarkonium requires opposite sign). Those terms with $1$ can be attributed to fluctuations feeding energy into the subsystem, the momentum dependent terms on the other hand encode dissipative effects that return energy to the medium. Together they allow the quarkonium particle to eventually thermalize with its surroundings. Both phenomena are governed by a function christened $D$, which is intimately related to the imaginary part of the complex EFT potential $D(r)={\rm Im}[V](r)-{\rm Im}[V](\infty)$. On the other hand, the Debye screened real part of the EFT potential enters in the modified Hamiltonian $\tilde H_{Q\bar{Q}}$ of the corresponding Lindblad equation \eqref{Eq:Lindblad}. Hence this QCD based result tells us how the complex heavy quark potential governs quarkonium dynamics at high temperatures, which clearly goes beyond a simple Schr\"odinger equation. The question of how to extend this approach, currently limited to high temperatures, to a strongly coupled medium is the focus of ongoing research efforts.

There is another bit of insight hiding in plain sight: the function $D(r)$ exhibits a well peaked structure around the origin $r=0$, which can be identified as a new intrinsic scale in the system, the correlation length $\ell_{\rm corr}$ of the medium (see e.g. discussion in Ref.\cite{Kajimoto:2017rel}). In the context of quarkonium dynamics it plays an important role, as it characterizes the effects of {\it decoherence} on the quarkonium wavefunction. If the $\ell_{\rm corr}$ is much larger than the extent of our quarkonium state, $L_{{\bf k},a}$ will induce essentially the same color rotations for the $Q$ and $\bar{Q}$ leaving the bound state almost unaffected. Once $\ell_{\rm corr}$ becomes similar to the quarkonium size, the quark and antiquark receive different color rotations and the binding of the state is strongly impaired. Why did we say that $\ell_{\rm corr}$ is related to decoherence though? When we have a closer look at the effect of the function $D$, we find that it acts approximately as $\rho_{Q\bar{Q}}({\bf x}_1,{\bf x}_2,t)\sim\rho({\bf x}_1,{\bf x}_2,0){\rm exp}\big[-\big(D(0)-D({\bf x}_1-{\bf x}_2)\big)t\big]$, i.e. it induces the decay of the off-diagonal terms of the density matrix, which is exactly what decoherence does. I.e. it is understood that the real-time dynamics of quarkonium arise from an intricate interplay of screening (characterized by the Debye mass $m_D$ ) and decoherence ( characterized by $\ell_{\rm corr}$ ).

If decoherence is efficient, one expects that a semiclassical description will eventually become applicable. As investigated in detail in \cite{Blaizot:2017ypk,Blaizot:2018oev}, if in addition the color degrees of freedom thermalize quickly, the non-Abelian nature of the theory does not play an important role and one may go over to a Langevin-like description of the heavy quarkonium states. One of the attractive features of such a description is the possibility to straight forwardly generalize it to many $Q\bar{Q}$ pairs, allowing insight into recombination dynamics which are difficult to compute in a fully quantum setting.

One aspect of the quarkonium dynamics that two poster contributions to this conference (\cite{Kajimoto:2020ip,Sharma:2019xum}) elucidated is the effect of decoherence in the presence of explicitly implemented color degrees of freedom. To this end both groups restricted themselves to the recoilless limit (the former in 1d, the latter in 3d), which is expected to be applicable at early times in the evolution. In this approximation the dynamics is represented by an ensemble of wavefunctions (that carry a color index) which are governed by a linear Schr\"odinger equation with real-valued attractive color singlet and repulsive octet potential, as well as two noise terms in color space. By inspecting the color singlet and octet density matrices, it was found that while decoherence is leading to a diagonal structure for large values of the separation distances, for small distances, remnants of the initial bound state may persist for an appreciable amount of time. Taking the Wigner transform of such a density matrix, the structures around the origin translate into negative contributions of the Wigner distribution function. This in turn signals that the genuine quantum nature of quarkonium can remain relevant during the evolution in a hot medium and that if we wish to apply a semi-classical approximation its applicability needs to be established on a case-by-case basis.

Another recent study presented as poster contribution at this conference \cite{Miura:2019ssi} took on the task to implement the fully dissipative dynamics of the Lindblad equation at high temperature discussed above. As is known in the condensed matter open quantum systems community, any Lindblad master equation can be implemented (the jargon term is unravelled) in terms of an ensemble of wavefunctions that evolve according to a non-linear stochastic Schr\"odinger equation. Using the unravelling technique of quantum state diffusion, the authors found that simulations, starting from different initial states (ground or first excited state), will at late enough times approach a universal steady state of non-vanishing occupation numbers (within the finite volume of a heavy-ion collision). This is encouraging, as both the deterministic Schr\"odinger equation with complex potential and the dissipationless stochastic potential model all lead to vanishing bound state contributions at late times. One may ask whether the steady state that is observed is a thermal state, which the authors answer in the affirmative by plotting the late-time occupancies vs. the mode energy. This revealed a clean exponential Boltzmann behavior, whose exponent excellently reproduced the environment temperature. The open quantum systems framework considered here is a versatile theoretical tool, as it provides a unified language to describe the evolution of a single heavy quark, a quarkonium pair and in principle also that of multiple $Q\bar{Q}$ pairs, even if the computational cost is extremely high. As a result e.g. the function $D(r)$, related to the imaginary part of the interquark potential in the medium also governs the dynamics of the single quark (see e.g. \cite{Akamatsu:2018xim}). A similar connection between quark transport and evolution of quarkonium we discussed in the context of the heavy quark diffusion coefficient and the pNRQCD based master equation.

Heavy quarkonium theory has entered a new and exciting era, in which the dissipative real-time evolution of quark-antiquark pairs takes center stage. Combining the open quantum systems framework with the arsenal of high precision lattice QCD simulations available, the community is working on non-perturbative implementations of quarkonium dynamics that connect QCD to actual phenomenology and experimental data with unprecendented carity. Together with the activities of the theory community in better understanding the initial production process (not to forget the advances in the lattice QCD determination of parton distribution functions), a more and more comprehensive picture of quarkonium production and suppression will emerge over the coming years. The prospect of the measurement of excited states in the upcoming run3 at LHC promises access to much more discriminatory observables, which together with the newly found insight on the range of validity of phenomenological models, will help us to distill the relevant physical processes underlying the measured yields. Exciting news about first measurements in heavy-ion collisions of exotic quarkonium states, such as the X(3872) presented by CMS at this conference \cite{YJL:2020umv} furthermore opens up a new venue to learn about the microscopic composition of these states, which have been studied by theorists at $T=0$ intensively. 

The author acknowledges funding by the Research Council of Norway under the FRIPRO Young Research Talent grant 286883 and grant 295310.\vspace{-0.3cm}


\bibliographystyle{elsarticle-num}
\bibliography{QM2019_Quarkonium_Rothkopf}

\begin{thebibliography}{10}
\expandafter\ifx\csname url\endcsname\relax
  \def\url#1{\texttt{#1}}\fi
\expandafter\ifx\csname urlprefix\endcsname\relax\def\urlprefix{URL }\fi
\expandafter\ifx\csname href\endcsname\relax
  \def\href#1#2{#2} \def\path#1{#1}\fi

\bibitem{Rothkopf:2019ipj}
A.~Rothkopf, {Heavy Quarkonium in Extreme Conditions}\href
  {http://arxiv.org/abs/1912.02253} {\path{arXiv:1912.02253}}.

\bibitem{Sirunyan:2017lzi}
A.~M. Sirunyan, et~al., {Suppression of Excited $\Upsilon$ States Relative to
  the Ground State in Pb-Pb Collisions at $\sqrt{s_\mathrm{NN}}$=5.02 TeV},
  Phys. Rev. Lett. 120~(14) (2018) 142301.

\bibitem{Acharya:2018pjd}
S.~Acharya, et~al., {Study of J/$\psi$ azimuthal anisotropy at forward rapidity
  in Pb-Pb collisions at $ \sqrt{s_{\mathrm{NN}}}=5.02 $ TeV}, JHEP 02 (2019)
  012.

\bibitem{Laine:2006ns}
M.~Laine, O.~Philipsen, P.~Romatschke, M.~Tassler, {Real-time static potential
  in hot QCD}, JHEP 03 (2007) 054.

\bibitem{Rothkopf:2011db}
A.~Rothkopf, T.~Hatsuda, S.~Sasaki, {Complex Heavy-Quark Potential at Finite
  Temperature from Lattice QCD}, Phys. Rev. Lett. 108 (2012) 162001.

\bibitem{Burnier:2014ssa}
Y.~Burnier, O.~Kaczmarek, A.~Rothkopf, {Static quark-antiquark potential in the
  quark-gluon plasma from lattice QCD}, Phys. Rev. Lett. 114~(8) (2015) 082001.

\bibitem{Petreczky:2018xuh}
P.~Petreczky, A.~Rothkopf, J.~Weber, {Realistic in-medium heavy-quark potential
  from high statistics lattice QCD simulations}, Nucl. Phys. A982 (2019) 735.

\bibitem{Kaczmarek:2020umv}
O.~Kaczmarek, L.~Altenkort, H.-T. Ding, A.-L. Kruse, H.~Ohno, S.~Hai-Tao,
  {Heavy quark diffusion coefficients and thermal quarkonium mass shifts from
  lattice QCD}, in: {28th International Conference on Ultrarelativistic
  Nucleus-Nucleus Collisions (Quark Matter 2019) Wuhan, China, November 4-9,
  2019}, 2020.

\bibitem{Brambilla:2019tpt}
N.~Brambilla, M.~A. Escobedo, A.~Vairo, P.~Vander~Griend, {Transport
  coefficients from in medium quarkonium dynamics}, Phys. Rev. D100~(5) (2019)
  054025.
\newblock \href {http://dx.doi.org/10.1103/PhysRevD.100.054025}
  {\path{doi:10.1103/PhysRevD.100.054025}}.

\bibitem{Cheung:2018upe}
V.~Cheung, R.~Vogt, {Production and polarization of prompt $\Upsilon$($n$S) in
  the improved color evaporation model using the $k_T$-factorization approach},
  Phys. Rev. D99~(3) (2019) 034007.
\newblock \href {http://dx.doi.org/10.1103/PhysRevD.99.034007}
  {\path{doi:10.1103/PhysRevD.99.034007}}.

\bibitem{Ma:2015sia}
Y.-Q. Ma, R.~Venugopalan, H.-F. Zhang, {$J/\psi$ production and suppression in
  high energy proton-nucleus collisions}, Phys. Rev. D92 (2015) 071901.
\newblock \href {http://dx.doi.org/10.1103/PhysRevD.92.071901}
  {\path{doi:10.1103/PhysRevD.92.071901}}.

\bibitem{Andronic:2019wva}
A.~Andronic, P.~Braun-Munzinger, M.~K. Köhler, K.~Redlich, J.~Stachel,
  {Transverse momentum distributions of charmonium states with the statistical
  hadronization model}, Phys. Lett. B797 (2019) 134836.
\newblock \href {http://dx.doi.org/10.1016/j.physletb.2019.134836}
  {\path{doi:10.1016/j.physletb.2019.134836}}.

\bibitem{He:2020umv}
M.~He, R.~Rapp, {Charm-hadron production in $pp$ and AA collisions}, in: {28th
  International Conference on Ultrarelativistic Nucleus-Nucleus Collisions
  (Quark Matter 2019) Wuhan, China, November 4-9, 2019}, 2020.
\newblock \href {http://arxiv.org/abs/2002.00392} {\path{arXiv:2002.00392}}.

\bibitem{Zhao:2020wpi}
W.~Zhao, C.~M. Ko, Y.-X. Liu, G.-Y. Qin, H.~Song, {Number of constituent quark
  scaling of elliptic flows in high multiplicity p-Pb collisions at
  $\sqrt{s_{NN}}=$ 5.02 TeV}, in: {28th International Conference on
  Ultrarelativistic Nucleus-Nucleus Collisions (Quark Matter 2019) Wuhan,
  China, November 4-9, 2019}, 2020.
\newblock \href {http://arxiv.org/abs/2001.10689} {\path{arXiv:2001.10689}}.

\bibitem{Matsui:1986dk}
T.~Matsui, H.~Satz, {$J/\psi$ Suppression by Quark-Gluon Plasma Formation},
  Phys. Lett. B178 (1986) 416--422.
\newblock \href {http://dx.doi.org/10.1016/0370-2693(86)91404-8}
  {\path{doi:10.1016/0370-2693(86)91404-8}}.

\bibitem{Asakawa:2000tr}
M.~Asakawa, T.~Hatsuda, Y.~Nakahara, {Maximum entropy analysis of the spectral
  functions in lattice QCD}, Prog. Part. Nucl. Phys. 46 (2001) 459--508.
\newblock \href {http://dx.doi.org/10.1016/S0146-6410(01)00150-8}
  {\path{doi:10.1016/S0146-6410(01)00150-8}}.

\bibitem{Burnier:2013nla}
Y.~Burnier, A.~Rothkopf, {Bayesian Approach to Spectral Function Reconstruction
  for Euclidean Quantum Field Theories}, Phys. Rev. Lett. 111 (2013) 182003.
\newblock \href {http://dx.doi.org/10.1103/PhysRevLett.111.182003}
  {\path{doi:10.1103/PhysRevLett.111.182003}}.

\bibitem{Aarts:2014cda}
G.~Aarts, C.~Allton, T.~Harris, S.~Kim, M.~P. Lombardo, S.~M. Ryan, J.-I.
  Skullerud, {The bottomonium spectrum at finite temperature from N$_{f}$ = 2 +
  1 lattice QCD}, JHEP 07 (2014) 097.
\newblock \href {http://dx.doi.org/10.1007/JHEP07(2014)097}
  {\path{doi:10.1007/JHEP07(2014)097}}.

\bibitem{Kim:2018yhk}
S.~Kim, P.~Petreczky, A.~Rothkopf, {Quarkonium in-medium properties from
  realistic lattice NRQCD}, JHEP 11 (2018) 088.
\newblock \href {http://dx.doi.org/10.1007/JHEP11(2018)088}
  {\path{doi:10.1007/JHEP11(2018)088}}.

\bibitem{Ikeda:2016czj}
A.~Ikeda, M.~Asakawa, M.~Kitazawa, {In-medium dispersion relations of charmonia
  studied by maximum entropy method}, Phys. Rev. D95~(1) (2017) 014504.
\newblock \href {http://dx.doi.org/10.1103/PhysRevD.95.014504}
  {\path{doi:10.1103/PhysRevD.95.014504}}.

\bibitem{Larsen:2020umv}
R.~Larsen, S.~Mukherjee, P.~Petreczky, S.~Meinel, {Bottomonia in QGP from
  lattice QCD: Beyond the ground states}, in: {28th International Conference on
  Ultrarelativistic Nucleus-Nucleus Collisions (Quark Matter 2019) Wuhan,
  China, November 4-9, 2019}, 2020.

\bibitem{Brambilla:2004jw}
N.~Brambilla, A.~Pineda, J.~Soto, A.~Vairo, {Effective Field Theories for Heavy
  Quarkonium}, Rev. Mod. Phys. 77 (2005) 1423.
\newblock \href {http://dx.doi.org/10.1103/RevModPhys.77.1423}
  {\path{doi:10.1103/RevModPhys.77.1423}}.

\bibitem{Beraudo:2007ky}
A.~Beraudo, J.~P. Blaizot, C.~Ratti, {Real and imaginary-time Q anti-Q
  correlators in a thermal medium}, Nucl. Phys. A806 (2008) 312--338.
\newblock \href {http://dx.doi.org/10.1016/j.nuclphysa.2008.03.001}
  {\path{doi:10.1016/j.nuclphysa.2008.03.001}}.

\bibitem{Brambilla:2008cx}
N.~Brambilla, J.~Ghiglieri, A.~Vairo, P.~Petreczky, {Static quark-antiquark
  pairs at finite temperature}, Phys. Rev. D78 (2008) 014017.
\newblock \href {http://dx.doi.org/10.1103/PhysRevD.78.014017}
  {\path{doi:10.1103/PhysRevD.78.014017}}.

\bibitem{Burnier:2012az}
Y.~Burnier, A.~Rothkopf, {Disentangling the timescales behind the
  non-perturbative heavy quark potential}, Phys. Rev. D86 (2012) 051503.
\newblock \href {http://dx.doi.org/10.1103/PhysRevD.86.051503}
  {\path{doi:10.1103/PhysRevD.86.051503}}.

\bibitem{Barnard:2019mnk}
N.~N. Barnard, W.~A. Horowitz, {Strongly coupled $\Upsilon$(1S) suppression in
  $\sqrt {s_{NN}}=$ 2.76 TeV Pb$+$Pb collisions}, J. Phys. Conf. Ser. 1271~(1)
  (2019) 012012.
\newblock \href {http://dx.doi.org/10.1088/1742-6596/1271/1/012012}
  {\path{doi:10.1088/1742-6596/1271/1/012012}}.

\bibitem{Liu:2017qah}
S.~Y.~F. Liu, R.~Rapp, {$T$-matrix Approach to Quark-Gluon Plasma}, Phys. Rev.
  C97~(3) (2018) 034918.
\newblock \href {http://dx.doi.org/10.1103/PhysRevC.97.034918}
  {\path{doi:10.1103/PhysRevC.97.034918}}.

\bibitem{Ferreiro:2012rq}
E.~G. Ferreiro, {Charmonium dissociation and recombination at LHC: Revisiting
  comovers}, Phys. Lett. B731 (2014) 57--63.
\newblock \href {http://dx.doi.org/10.1016/j.physletb.2014.02.011}
  {\path{doi:10.1016/j.physletb.2014.02.011}}.

\bibitem{Zhao:2011cv}
X.~Zhao, R.~Rapp, {Medium Modifications and Production of Charmonia at LHC},
  Nucl. Phys. A859 (2011) 114--125.
\newblock \href {http://dx.doi.org/10.1016/j.nuclphysa.2011.05.001}
  {\path{doi:10.1016/j.nuclphysa.2011.05.001}}.

\bibitem{Liu:2009nb}
Y.-p. Liu, Z.~Qu, N.~Xu, P.-f. Zhuang, {J/psi Transverse Momentum Distribution
  in High Energy Nuclear Collisions at RHIC}, Phys. Lett. B678 (2009) 72--76.
\newblock \href {http://dx.doi.org/10.1016/j.physletb.2009.06.006}
  {\path{doi:10.1016/j.physletb.2009.06.006}}.

\bibitem{Krouppa:2016jcl}
B.~Krouppa, M.~Strickland, {Predictions for bottomonia suppression in 5.023 TeV
  Pb-Pb collisions}, Universe 2~(3) (2016) 16.
\newblock \href {http://dx.doi.org/10.3390/universe2030016}
  {\path{doi:10.3390/universe2030016}}.

\bibitem{Krouppa:2017jlg}
B.~Krouppa, A.~Rothkopf, M.~Strickland, {Bottomonium suppression using a
  lattice QCD vetted potential}, Phys. Rev. D97~(1) (2018) 016017.
\newblock \href {http://dx.doi.org/10.1103/PhysRevD.97.016017}
  {\path{doi:10.1103/PhysRevD.97.016017}}.

\bibitem{Yao:2020kqy}
X.~Yao, W.~Ke, Y.~Xu, S.~A. Bass, T.~Mehen, B.~Müller, {Quarkonium Production
  in Heavy Ion Collisions: From Open Quantum System to Transport Equation},
  2020.
\newblock \href {http://arxiv.org/abs/2002.04079} {\path{arXiv:2002.04079}}.

\bibitem{Yao:2018nmy}
X.~Yao, T.~Mehen, {Quarkonium in-medium transport equation derived from first
  principles}, Phys. Rev. D99~(9) (2019) 096028.
\newblock \href {http://dx.doi.org/10.1103/PhysRevD.99.096028}
  {\path{doi:10.1103/PhysRevD.99.096028}}.

\bibitem{Yao:2018sgn}
X.~Yao, B.~Müller, {Quarkonium inside the quark-gluon plasma: Diffusion,
  dissociation, recombination, and energy loss}, Phys. Rev. D100~(1) (2019)
  014008.
\newblock \href {http://dx.doi.org/10.1103/PhysRevD.100.014008}
  {\path{doi:10.1103/PhysRevD.100.014008}}.

\bibitem{Katz:2015qja}
R.~Katz, P.~B. Gossiaux, {The Schrödinger–Langevin equation with and without
  thermal fluctuations}, Annals Phys. 368 (2016) 267--295.
\newblock \href {http://arxiv.org/abs/1504.08087} {\path{arXiv:1504.08087}},
  \href {http://dx.doi.org/10.1016/j.aop.2016.02.005}
  {\path{doi:10.1016/j.aop.2016.02.005}}.

\bibitem{Akamatsu:2011se}
Y.~Akamatsu, A.~Rothkopf, {Stochastic potential and quantum decoherence of
  heavy quarkonium in the quark-gluon plasma}, Phys. Rev. D85 (2012) 105011.
\newblock \href {http://dx.doi.org/10.1103/PhysRevD.85.105011}
  {\path{doi:10.1103/PhysRevD.85.105011}}.

\bibitem{Akamatsu:2012vt}
Y.~Akamatsu, {Real-time quantum dynamics of heavy quark systems at high
  temperature}, Phys. Rev. D87~(4) (2013) 045016.
\newblock \href {http://dx.doi.org/10.1103/PhysRevD.87.045016}
  {\path{doi:10.1103/PhysRevD.87.045016}}.

\bibitem{Akamatsu:2014qsa}
Y.~Akamatsu, {Heavy quark master equations in the Lindblad form at high
  temperatures}, Phys. Rev. D91~(5) (2015) 056002.
\newblock \href {http://dx.doi.org/10.1103/PhysRevD.91.056002}
  {\path{doi:10.1103/PhysRevD.91.056002}}.

\bibitem{Kajimoto:2017rel}
S.~Kajimoto, Y.~Akamatsu, M.~Asakawa, A.~Rothkopf, {Dynamical dissociation of
  quarkonia by wave function decoherence}, Phys. Rev. D97~(1) (2018) 014003.
\newblock \href {http://dx.doi.org/10.1103/PhysRevD.97.014003}
  {\path{doi:10.1103/PhysRevD.97.014003}}.

\bibitem{Blaizot:2017ypk}
J.-P. Blaizot, M.~A. Escobedo, {Quantum and classical dynamics of heavy quarks
  in a quark-gluon plasma}, JHEP 06 (2018) 034.
\newblock \href {http://dx.doi.org/10.1007/JHEP06(2018)034}
  {\path{doi:10.1007/JHEP06(2018)034}}.

\bibitem{Blaizot:2018oev}
J.-P. Blaizot, M.~A. Escobedo, {Approach to equilibrium of a quarkonium in a
  quark-gluon plasma}, Phys. Rev. D98~(7) (2018) 074007.
\newblock \href {http://dx.doi.org/10.1103/PhysRevD.98.074007}
  {\path{doi:10.1103/PhysRevD.98.074007}}.

\bibitem{Kajimoto:2020ip}
S.~Kajimoto, Y.~Akamatsu, M.~Asakawa, A.~Rothkopf, (in progress).

\bibitem{Sharma:2019xum}
R.~Sharma, A.~Tiwari, {Quantum evolution of quarkonia with correlated and
  uncorrelated noise}\href {http://arxiv.org/abs/1912.07036}
  {\path{arXiv:1912.07036}}.

\bibitem{Miura:2019ssi}
T.~Miura, Y.~Akamatsu, M.~Asakawa, A.~Rothkopf, {Quantum Brownian motion of a
  heavy quark pair in the quark-gluon plasma}\href
  {http://arxiv.org/abs/1908.06293} {\path{arXiv:1908.06293}}.

\bibitem{Akamatsu:2018xim}
Y.~Akamatsu, M.~Asakawa, S.~Kajimoto, A.~Rothkopf, {Quantum dissipation of a
  heavy quark from a nonlinear stochastic Schr\"odinger equation}, JHEP 07
  (2018) 029.
\newblock \href {http://dx.doi.org/10.1007/JHEP07(2018)029}
  {\path{doi:10.1007/JHEP07(2018)029}}.

\bibitem{YJL:2020umv}
Y.-J.~L. for~the [CMS~Collaboration], {Observation of X(3872) in PbPb
  collisions with the CMS detector}, in: {28th International Conference on
  Ultrarelativistic Nucleus-Nucleus Collisions (Quark Matter 2019) Wuhan,
  China, November 4-9, 2019}, 2020.

\end{thebibliography}







\end{document}